\documentclass[a4paper]{article}

\usepackage{INTERSPEECH2020}

\usepackage{xcolor,enumitem}
\usepackage{amsmath,amssymb,amsfonts,amsthm}
\usepackage{url}

\newcommand{\laurent}[1]{\textcolor{black}{#1}}
\newcommand{\thomas}[1]{\textcolor{black}{#1}}

\title{What the Future Brings: Investigating the Impact of Lookahead \\ for Incremental Neural TTS}
\name{Brooke Stephenson$^1$$^,$$^2$, Laurent Besacier$^2$, Laurent Girin$^1$, Thomas Hueber$^1$}
\address{
  $^1$Université Grenoble Alpes, CNRS, Grenoble INP, GIPSA-lab, 38000 Grenoble, France\\
  $^2$LIG, UGA, G-INP, CNRS, INRIA, Grenoble, France}
\email{brooke.stephenson@gipsa-lab.grenoble-inp.fr, 
laurent.besacier@univ-grenoble-alpes.fr, laurent.girin@gipsa-lab.grenoble-inp.fr, thomas.hueber@gipsa-lab.grenoble-inp.fr}

\begin{document}

\maketitle
\begin{abstract}
In incremental text to speech synthesis (iTTS), the synthesizer produces an audio output before it has access to the entire input sentence. In this paper, we study the behavior of a neural sequence-to-sequence TTS system when used in an incremental mode, i.e. when generating speech output for token $n$, the system has access to $n+k$ tokens from the text sequence. We first analyze the impact of this incremental policy on the evolution of the encoder representations of token $n$ for different values of $k$ (the lookahead parameter). The results show that, on average, tokens travel $88\%$ of the way to their full context representation with a one-word lookahead and $94\%$ after 2 words. We then investigate which text features are the most influential on the evolution towards the final representation using a random forest analysis. The results show that the most salient factors are related to token length. We finally evaluate the effects of lookahead $k$ at the decoder level, using a MUSHRA listening test. This test shows results that contrast with the above high figures: speech synthesis quality obtained with 2 word-lookahead is significantly lower than the one obtained with the full sentence.

\end{abstract}

\section{Introduction}
Text-to-speech (TTS) systems have made great strides with the introduction of sequence-to-sequence (seq2seq) neural models, combined with end-to-end trainable architectures \cite{shen2018natural,arik2017deep,sotelo2017char2wav,Wang2017}. Neural models typically take character as input and learn a direct mapping to spectrogram or waveform output, without the need for feature engineering. 
However, most of these neural TTS systems are designed to work at the sentence level, i.e. the synthetic speech signal is generated after the user has typed a complete sentence. When processing a given word, the system can thus rely on its full linguistic context (i.e. both past and future words) to build its internal representation.
 Despite its ability to generate high-quality speech, this synthesis paradigm is not ideal for several applications. For example, when used as a substitute voice by people with severe communication disorders or integrated in a dialog system (e.g. personal assistant, simultaneous speech interpretation, etc.), the system's need to wait until the end
 of a sentence introduces a latency which might be disruptive to conversational flow and system interactivity. 
Incremental TTS (iTTS, sometimes called low-latency or online TTS) aims to address these issues by synthesizing speech on-the-fly, that is by outputting audio chunks as soon as a new word (or a few of them) become available. This task is particularly challenging since producing speech without relying on the full linguistic context can result in both segmental (phonetic) and supra-segmental (prosodic) errors \cite{lemaguer2013}. 

Early iTTS systems were developed in the context of HMM-based speech synthesis \cite{Baumann2012,pouget2015hmm,pouget_2016_interspeech_adaptive_latency_iTTS}. In this paradigm, models are trained on a set of explicit linguistic features (e.g. number of syllables in the next word). The authors of \cite{Baumann2012,pouget2015hmm} developed coping mechanisms to handle missing features when making predictions for iTTS: unknown future context information is replaced with the most common values for these features at inference time in \cite{Baumann2012}, whereas uncertainty on those features is explicitly integrated at training time by \cite{pouget2015hmm}. In \cite{pouget_2016_interspeech_adaptive_latency_iTTS}, an adaptive decoding policy based on the online estimation of the stability of the linguistic features is proposed: the synthesis of a given word is delayed if its part-of-speech (POS) is likely to change when additional (future) words are added.  

Several strategies have been proposed to reduce the latency of a sequence-to-sequence model with input text for neural machine translation \cite{Ma19acl, Arivazhagan19acl, Ma20iclr} or incremental speech translation~\cite{DBLP:journals/corr/abs-1808-00491,oda2014optimizing,gu-etal-2017-learning}. However, only a few studies have attempted to adapt these models for iTTS \cite{Yanagita2019, ma2019incremental}. The authors of \cite{Yanagita2019} proposed an approach that consists in (1) marking three subunits within the training sentences using \textit{start}, \textit{middle} and \textit{end} tags, (2) training a Tacotron 2 TTS model with these tags so it learns intrasentential boundary characteristics, and  (3) synthesizing sentences by inputting chunks of length $n$ words (up to half a sentence) with the appropriate \textit{middle} or \textit{end} tag. An alternate policy  reported in \cite{ma2019incremental} \laurent{(inspired by the prefix-to-prefix framework introduced for  translation \cite{Ma19acl}) consists in having access to a future context of $k$ input tokens while generating speech output. They also rely on the soft attention to learn the relationship between the predicted spectrogram and the currently available source text.}
These two approaches give promising results but introduce a fixed size (and possibly large) latency.  

The goal of the present paper is to pave the way toward an adaptive decoding policy for a neural iTTS. Similarly to the HMM-based iTTS system described in \cite{pouget_2016_interspeech_adaptive_latency_iTTS}, the envisioned neural iTTS is expected to modulate the lookahead (and thus the latency) by the uncertainty on some features due to the lack of future context. However, the gain in naturalness provided by end-to-end models (over, e.g., HMM-based systems) is also accompanied by reduced interpretability. Because of the black box nature of the models, studying the importance of missing features is a challenging task.
To address this, 
we analyse the evolution of the encoder representations of a neural TTS (Tacotron 2) when words are incrementally added (i.e. when generating speech output for token $n$, the system only has access to $n+k$ tokens from the text sequence, $k$ being the lookahead parameter). We also investigate which text features are the most influential on this evolution towards the final encoder representation. Finally, we evaluate the effect of the lookahead at the perceptual level using a MUSHRA listening test.

\section{Methods and Materials}
\label{method}

\subsection{Models and data}

For these experiments, we use a sequence-to-sequence neural model which has achieved state-of-the-art results: Tacotron 2 \cite{shen2018natural}. 
We use a pretrained model developed by NVIDIA\footnote{\url{https://github.com/NVIDIA/tacotron2}} and trained on the LJ Speechset \cite{Ito2017}, a collection of non-fiction books read by a single speaker. The corpus contains 24 hours of audio recordings.
The encoder takes characters as input, passes them through an embedding layer, three convolution layers and then a bidirectional LSTM layer. The decoder uses the encoder output, an attention module and previous decoder outputs to predict the corresponding log-spectrogram frames, which are converted into a speech waveform using \thomas{WaveGlow neural vocoder} 
\cite{prenger2019waveglow}.

For our analysis, the \thomas{test} sentences used as input sequences are taken from the libriTTS corpus \cite{zen2019libritts}. We filter $1,000$ utterances with sentence length ranging from $5$ to $42$ words. We follow the procedure outlined in \cite{kilgarriff2001comparing} to verify that word distribution is similar to that of larger general corpora.\footnote{Brown and BNC corpora.} Our corpus contains $34,768$ tokens and $4,085$ types.

\subsection{Incremental encoding policy}
\label{policy}

We consider an input sequence of tokens, where each token can be either a word, a space or a punctuation mark.
We define an iTTS system with the following simple policy \laurent{(similar to \cite{ma2019incremental})}: the encoder starts by reading $k$ input tokens ($k$ is the lookahead parameter) and then it alternates between generating speech output and reading the next token until the complete input token sequence is consumed.
Formally, we use the following notations and definitions (see Table \ref{tab:increments} for an example of the listed items):
\begin{itemize}[leftmargin=*]
\item $N$ is the length of the input sequence (in number of tokens); 
\item $\mathbf{x}_n$ is the token at position $n$ (the ``current'' token); $\mathbf{x}_{1:N}$ is the complete sequence of input tokens; $\mathbf{x}_{1:n}$ is the subsequence of input tokens from position $1$ to position $n$;
\item $\mathbf{y}_n$ is the speech output segment corresponding to token  $\mathbf{x}_n$;
\item $c(n,k) = \min(n+k,N)$ is the number of input tokens read when generating $\mathbf{y}_n$ (recall that $k$ is the lookahead parameter); $\mathbf{z}^{n,k}_{n}$ is the corresponding encoder output.\footnote{$n$ is used two times in $\mathbf{z}^{n,k}_{n}$ since we will see that the value at other positions, e.g. $\mathbf{z}^{n,k}_{n-1}$, also depends on $n$ and $k$.} In other words, $\mathbf{y}_n$ is obtained after reading the partial sequence of input tokens $\mathbf{x}_{1:c(n,k)}$; $\mathbf{z}^{n,k}_{1:c(n,k)}$ is the sequence of encoder representations obtained so far;
\end{itemize}
Conventional offline encoding (using the full sequence of input tokens $\mathbf{x}_{1:N}$ at each position $n$) is also processed for comparison, and $\mathbf{z}_{1:N}^{\text{full}}$ denotes the corresponding encoded sequence. 

\begin{table}[ht!]
\caption{Incremental inputs (for different lookahead $k$) for sentence ``The dog is in the yard.'' to generate $\mathbf{x}_3$ (the word ``dog.'')}
\label{tab:increments}
\centering
\begin{tabular}{|l|l|l|l|l|}
\hline
$n$ & $k$ & $c(n,k)$ & Input at $c(n,k)$      & $\mathbf{x}_n$ \\ \hline
3            & 0            & 3              & The\_dog                     & dog            \\ \hline
3            & 1            & 4              & The\_dog\_                   & dog            \\ \hline
3            & 2            & 5              & The\_dog\_is                 & dog            \\ \hline
3            & ...            & ...              & ...               & dog            \\ \hline
3            & 8            & 11             & The\_dog\_is\_in\_the\_yard  & dog            \\ \hline
3            & 9            & {$N=12$}            & The\_dog\_is\_in\_the\_yard. & dog            \\ \hline
\end{tabular}
  \label{tab:exemple}
\end{table}


\subsection{From character to word representations}

\begin{figure}[ht!]
  \centering
  \includegraphics[width=\linewidth]{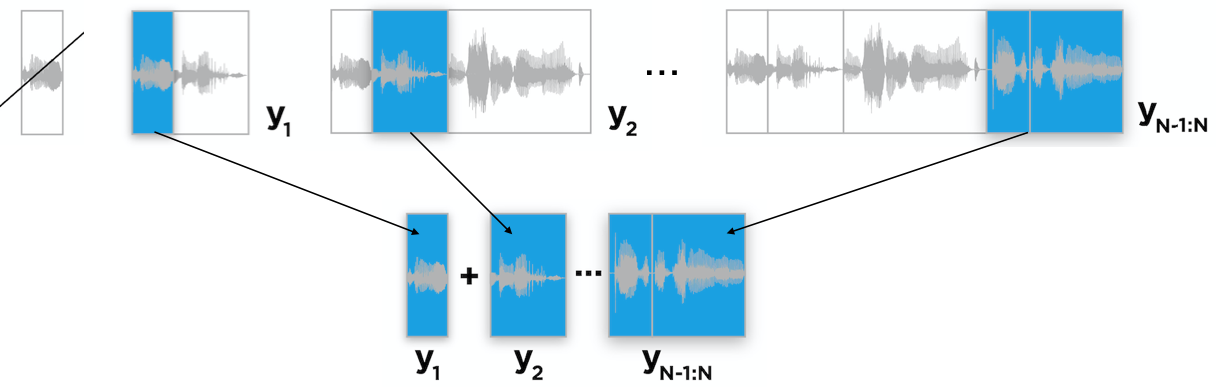}
  \caption{Illustration of the incremental speech waveform generation process for lookahead parameter $k=1$.} 
  \label{fig:concat}
\end{figure}

In the Tacotron model, input sequences are encoded at the character level. However, in our study, we consider an iTTS decoding policy at the word level; this is because token breaks are a natural trigger for synthesis or evaluation in a practical iTTS system. Consequently, we need to go from character representation to word representation.
We start from the encoder's bidirectional LSTM network: forward and backward layers each provide a 256-dimensional vector for each input character.
For each new token $\mathbf{x}_n$, we extract the output of the forward layer corresponding to the last character of $\mathbf{x}_n$. The forward layer continues up to the last character of token $\mathbf{x}_{c(n,k)}$. Then the backward layer goes from the last character of token $\mathbf{x}_{c(n,k)}$ to the first character of token $\mathbf{x}_{1}$. We extract the output of the backward layer corresponding to the first character of $\mathbf{x}_{n}$. Both vectors are concatenated to get a 512-dimensional vector representation $\mathbf{z}_n
^{n,k}$ of $\mathbf{x}_n$. Note that the input  sequence is re-encoded for each new token (i.e., for each increment of $n$), leading to new values for the sequence $\mathbf{z}_{1:n-1}^{n,k}$.
Of course, this sequence also depends on $k$, which is the purpose of this study. 
In contrast, there is only one single value for the sequence $\mathbf{z}_{1:N}^{\text{full}}$. 

\subsection{Incremental decoding} 

We build the iTTS decoder output as follows.  
For a given value of $k$, and for the current token $\mathbf{x}_{n}$, we first produce the speech waveform corresponding to the encoded sequence $\mathbf{z}_{1:n}^{n,k}$. 
Then, using the Munich Automatic Segmentation system \cite{kisler2017multilingual} (an automatic speech recognition and forced alignment tool which employs an HMM and Viterbi decoding to find the best alignment between the text and audio), we select the portion $\mathbf{y}_{n}$ of the waveform corresponding to $\mathbf{x}_{n}$. Finally we concatenate this speech segment $\mathbf{y}_{n}$ to the speech segment resulting from the processing of previous tokens, that we can denote as $\mathbf{y}_{1:n-1}$. In short, we simply update the generated speech waveform as $\mathbf{y}_{1:n} = [\mathbf{y}_{1:n-1} \ \mathbf{y}_{n}]$.
For example, for $k=2$, we extract the speech waveform segment $\mathbf{y}_1$ corresponding to token $\mathbf{x}_1$ from the signal generated from $\mathbf{z}_{1:3}^{1,2}$; then we extract the speech waveform segment $\mathbf{y}_2$ corresponding to token $\mathbf{x}_2$ from the signal generated from $\mathbf{z}_{1:4}^{1,2}$; we concatenate $\mathbf{y}_1$ and $\mathbf{y}_2$, and we continue this process until the end of the input sequence is read. This process is illustrated in Fig.~\ref{fig:concat} for $k=1$.
Segment concatenation is done with a 5-ms cross-fade, a simple and efficient way to prevent audible artefacts in our experiments. 
Note that the overall encoding and decoding process simulates an effective $k$-lookahead iTTS system that generates a new speech segment $\mathbf{y}_{n}$ when entering the new input token $\mathbf{x}_{c(n,k)}$. Sound examples obtained with this procedure are available online.\footnote {\url{https://tinyurl.com/y3hvl5cn}}

\begin{figure}[ht]
  \centering
  \includegraphics[width=\linewidth]{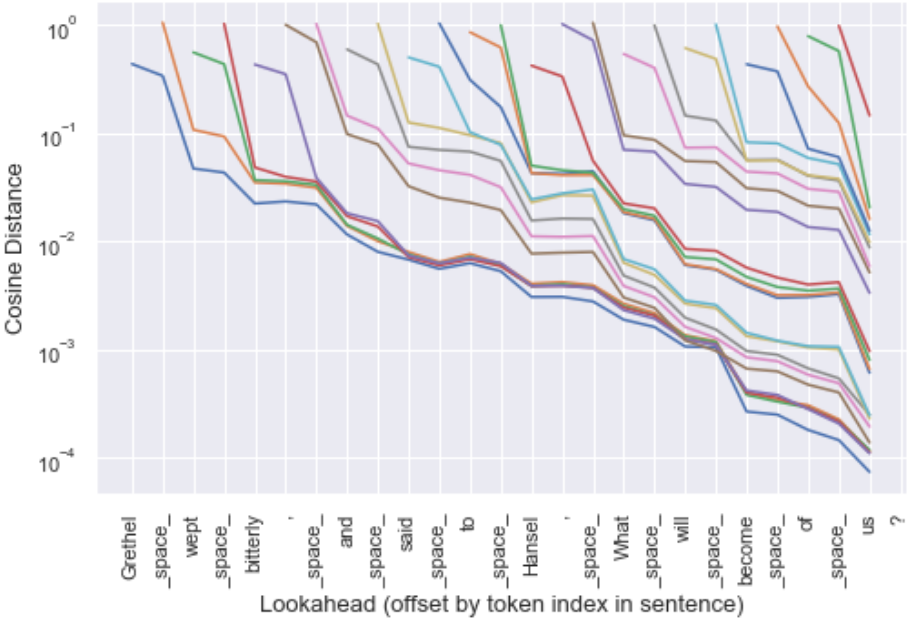}
  \caption{
  Change in token representations over time. Each colored line represents a token from the utterance “Grethel wept bitterly, and said to Hansel, What will become of us?” The height of the colored line shows the distance between encoder outputs $\mathbf{z}_{n}^{n,k}$ (incremental decoding) and $\mathbf{z}_{n}^{\text{full}}$ (offline decoding) at different values of k. The vertical grid line where $\mathbf{x}_n$ (the colored line) first appears represents $k=0$ for that token; the next vertical grid line to the right represents $k=1$ for $\mathbf{x}_n$ and $k=0$ for $\mathbf{x}_{n+1}$, etc.}
  \label{fig:overTime}
\end{figure}

\subsection{Analyzing   impact of lookahead  on encoder representation}
\label{encoderanalysis}



Our first goal is to analyze the impact of the lookahead parameter $k$ on the representation of a given token $\mathbf{x}_n$ at the encoder level. 
Given the two encoder representations of $\mathbf{x}_n$ ($\mathbf{z}_{n}^{n,k}$ in incremental mode and $\mathbf{z}_{n}^{\text{full}}$ in offline mode), we compute the cosine distance between them as 
$d(n,k)=1-\frac{\mathbf{z}_{n}^{n,k}.\mathbf{z}_{n}^{\text{full}}}{||\mathbf{z}_{n}^{n,k}||.||\mathbf{z}_{n}^{\text{full}}||}$.
We then average this distance for all tokens of our corpus or all tokens of a given syntactic category.
Our analysis consists in investigating which token features could best explain the observed variance in our data (i.e. why are some tokens relatively far from their final representation while others are close at the same value of $k$?). 
We did this using random forest (RF) regressors \cite{pedregosa2011scikit} which optimize cosine distance predictions and can provide information about which input features contribute the most towards these predictions.  
Our selected features are summarized in Table \ref{tab:features}.
The RFs were fit using 100 estimators, mean squared error measures and bootstrapping.
We followed the following procedure to determine which features are the most significant: (1) we add a column of random variables to our data set; (2) we fit an initial RF and eliminate all variables with a Gini importance lower than the random feature; and (3) we fit a new RF using only the remaining features and then calculate the permutation feature importance (i.e. the drop in $R^2$ that results from swapping columns in the dataset) \cite{altmann2010permutation}.

\begin{figure}[ht!]
  \centering
  \includegraphics[width=\linewidth]{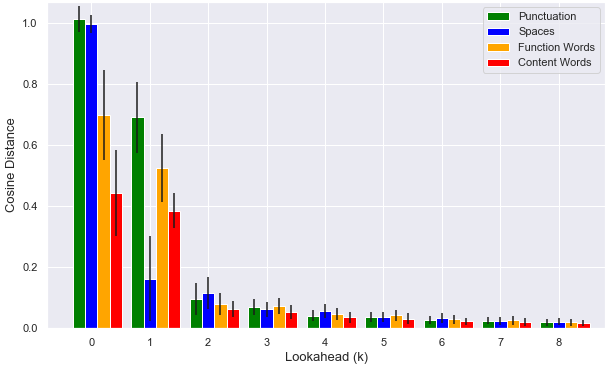}
  \caption{Distance $d(n,k)$ between 
  encoder representations $\mathbf{z}_{n}^{\text{full}}$ (offline) and $\mathbf{z}_{n}^{n,k}$ (incremental decoding) averaged over all tokens of a given category (punctuation, space, function word, content word), for lookahead parameter $k=0$ to $8$. Error bars represent standard deviation.} 
  \label{fig:POS}
\end{figure}

\subsection{Analyzing the effect of lookahead on decoder output}

We evaluate the perceptual impact of the lookahead $k$ using a MUSHRA listening test \cite{mushra}. To that purpose, we selected 20 sentences and generated each at multiple values of $k$, namely $k=1,2,4,6$. $k=1$ corresponds to a lookahead of one space (or one punctuation mark). It was chosen as the baseline and should be considered as the low-range anchor for the test. In general, $k=2$ represents a 1-word lookahead, $k=4$ a 2-word lookahead and $k=6$ a 3-word lookahead, although other cases occasionally happen (e.g. a space followed by an open parenthesis). Note that $k=0$ was not selected because the output signal was deemed too unintelligible to warrant evaluation. The reference stimuli were generated with the offline TTS mode, and were used both as reference and as the hidden high-range anchor. 21 participants, all native English speakers, were asked to assess the similarity between the reference and each of the stimuli obtained with the incremental decoding policy (plus the high-range anchor) on a 0-100 scale (100 means that sample and reference are identical). The MUSHRA test was done online, using the Web Audio Evaluation Tool \cite{waet2015}. 
3 participants were excluded from  analysis because they did not give high similarity ratings for the reference and the hidden high-range anchor (which were identical). Statistical significance between different experimental conditions (different values of $k$ and incremental vs. offline synthesis) were assessed using paired t-tests.

\section{Results and Discussion}
\label{results}

\subsection{Encoder representations}
\label{sec:res_enc_rep}
Figure~\ref{fig:overTime} displays an example of distance (in log-scale) between encoder representations in incremental \textit{versus} offline modes for a given sentence and all possible values of $k$.
While the global trend is a movement towards the final (offline) representation as $k$ increases ($d(n,k)$ decreases with $k$), we also observe some cases where an increment of the context leads to a representation that is farther away from its offline counterpart (see for instance the comma after the word ``Hansel''). 
One possible explanation for this might be that Tacotron interprets the input as the end of an intonational or rhythmic phrase and when further input is received, it reassesses the \laurent{token representation}.

\thomas{Figure~\ref{fig:POS} also shows the distance $d(n,k)$ for $k=0$ to $8$, but this time averaged over all tokens of the $1,000$ test sentences and for the different token categories: punctuation, space, function word and content word.}
\thomas{As in Figure~\ref{fig:overTime}, increasing the lookahead consistently reduces the distance between the encoder outputs in incremental and in offline mode, on average. Importantly, the most significant decrease is observed between $k=1$ and $k=2$, that is, when considering a lookahead of one space and one word (in addition to the current word). A slower decrease toward the final representation is observed for $k \geq 2$. This is consistent with Figure~\ref{fig:overTime}. A series of paired t-tests on $d(n,k)$ (averaged over all test sentences and all token categories) reveals a tiny but systematically significant difference between pairs of consecutive lookaheads (e.g. $k=3$ vs. $k=4$, $k=7$ vs. $k=8$) up to the end of the sentence. This might show that, on average, each new token considered in future context contributes slightly but significantly to the evolution of the encoder representation.}
We also observe that, while representations of content words are more stable to context variation, those of punctuation, spaces and function words are further away from their final representation in offline mode when not enough context is given ($k<2$). 

A more fine grain analysis of the factors that impact $d(n,k)$ is provided by
the results of the RF analysis, which are summarized in Table \ref{tab:features}. For $k=0$, the length of $\mathbf{x}_n$ is the most effective predictor of cosine distance, and for $k=2$ the lengths of $\mathbf{x}_{n+1}$ and $\mathbf{x}_{n+2}$ (i.e. the future tokens that the encoder sees when encoding $\mathbf{x}_n$) are the most effective predictors. 
\laurent{For instance,} at $k=2$, our model correctly predicts that the token ``to'' in Sentence A below (lookahead = \emph{space} + ``be'') is farther away from its final representation than ``to'' in Sentence B (lookahead = \emph{space} + ``Kitty''). The cosine distances are 0.135 and 0.057 respectively. 
\begin{itemize}[label={A)}]
\item{\textit{I suppose, he said, I ought \textbf{to} be glad of that.}}
\end{itemize}
\begin{itemize}[label={B)}]
\item {\textit{And the Captain of course concluded (after having been introduced \textbf{to} Kitty) that Mrs Norman was a widow.}}
\end{itemize}


\begin{table}[]
\caption{Influence of text features on the distance $d(n,k)$ 
estimated by RF regression for $k=0$ and $2$ (NS=not significant; 
*~=~weak effect; ** = medium effect; *** = strong effect).} 
\centering
\fontsize{8}{8}\selectfont
\begin{tabular}{|l|l|l|l|}
\hline
 &                                                                               &\multicolumn{2}{l|}{\begin{tabular}[c]{@{}l@{}}Permutation Feature \\ Importance\end{tabular}} \\ \hline
\textbf{Feature}                                                        & \textbf{Definition}                                                                                          & \textbf{$k=0$}          & \textbf{$k=2$}                                                                       \\ \hline
\begin{tabular}[c]{@{}l@{}}Token Length\end{tabular}                  & \begin{tabular}[c]{@{}l@{}}\# of characters in\\ $x_n$\end{tabular}                                       & ***                  & **                                                                               \\ \hline
POS                                                                     & \begin{tabular}[c]{@{}l@{}}Part of speech \\of $x_n$\end{tabular}                                            & NS                   & NS                                                                                 \\ \hline
\begin{tabular}[c]{@{}l@{}}Frequency \\ in Training\end{tabular}         & \begin{tabular}[c]{@{}l@{}}\# of instances of \\ $x_n$ in LJ Speechset\end{tabular}                     & *                  & NS                                                                                   \\ \hline
\begin{tabular}[c]{@{}l@{}}Relative \\ Position\end{tabular}                                                        & \begin{tabular}[c]{@{}l@{}}Token's relative \\ position in input \\ sequence = $n/N$\end{tabular}      & *                      & *                                                                               \\ \hline
Penultimate                                                             & Does $n = N-1$?                                                                                            & *                  & NS                                                                                   \\ \hline
\begin{tabular}[c]{@{}l@{}}Followed by \\ Punctuation\end{tabular}      & \begin{tabular}[c]{@{}l@{}}Is $x_{n+1}$ a punc-\\ tuation mark?\end{tabular}                                  & NS                      & NS                                                                                    \\ \hline
\begin{tabular}[c]{@{}l@{}}Distance to\\ Punctuation\end{tabular}       & \begin{tabular}[c]{@{}l@{}}\# of tokens before \\ next punctuation \\mark\end{tabular}                  & *                        & *                                                                               \\ \hline
\begin{tabular}[c]{@{}l@{}}Distance to\\ Parent Phrase \\End\end{tabular} & \begin{tabular}[c]{@{}l@{}}\# of tokens to the \\ end of parent cons- \\tituent group of $x_n$\end{tabular} & NS                      & NS                                                                                   \\ \hline
POSPrev + $m$                                                             & \begin{tabular}[c]{@{}l@{}}Part of speech of \\ token $x_{n-m}$\end{tabular}                                 & NS                      & NS                                                                          \\ \hline
POSNext + $m$                                                             & \begin{tabular}[c]{@{}l@{}}Part of speech of \\ token $x_{n+m}$\end{tabular}                                 & NS                      & $m$=1 *                                                                                   \\ \hline
\begin{tabular}[c]{@{}l@{}}Word Length\\ of Prev + $m$\end{tabular}      & \begin{tabular}[c]{@{}l@{}}\# of characters in \\ $x_{n-m}$\end{tabular}                                     & \begin{tabular}[c]{@{}l@{}}$m=1$ * \\ $m=2$ * \end{tabular}                       &    NS                                                                                \\ \hline
\begin{tabular}[c]{@{}l@{}}Word Length \\ of Next + $m$\end{tabular}       & \begin{tabular}[c]{@{}l@{}}\# of characters in\\ $x_{n+m}$\end{tabular}                                      &   $m=2$  *                    & \begin{tabular}[c]{@{}l@{}}$m=1$: ***\\ $m=2$: ***\\ $m=4$: *\end{tabular}          \\ \hline
\end{tabular}
\label{tab:features}
\end{table}

\begin{figure}[ht]
  \centering
  \includegraphics[width=\linewidth]{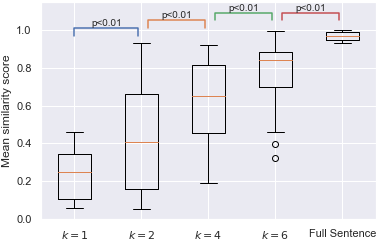}
  \caption{Perceptual evaluation of the impact of lookahead parameter $k$ using MUSHRA listening test.
  }
  \label{fig:mushra}
\end{figure}

\subsection{Perceptual evaluation of the decoder output}
\thomas{Results of the MUSHRA listening test are presented in Figure~\ref{fig:mushra}. First, statistical analyses show significant differences for all pairs of considered lookahead ($k=1$ vs. $k=2$, $k=2$ vs. $k=4$, and $k=4$ vs. $k=6$). This confirms at a perceptual level the tendency observed on the evolution of the encoder representation (see Section \ref{sec:res_enc_rep}): each additional lookahead brings the incremental synthesis closer to the offline one. We also found a significant difference between $k=6$ and $k=N$ (offline mode), i.e. with a lookahead of typically 3 words. This is in contradiction with  \cite{Yanagita2019} who did not report any difference between incremental and offline synthesis for such lookahead. Possible explanations for this include 1) the use of a different experimental paradigm (MUSHRA vs. MOS in \cite{Yanagita2019}), 2) duration distortions caused by the concatenation of speech segments or 3) by the fact that contrary to \cite{Yanagita2019}, we did not retrain the Tacotron 2 on shorter linguistic units (this is left for future work).}

\vspace{-0.2cm}

\section{Conclusion}
\label{conclusion}

This study presents several experiments which probe the impact of future context in a neural TTS system, based on a sequence-to-sequence model, both in terms of encoder representation and perceptual effect. Reported experimental results allow us to draw the contours of an adaptive decoding policy for an incremental neural TTS, which modulates the lookahead (and thus the overall latency) by potential change in internal representations. 
Shorter words are more dependent on future context than longer ones. Therefore, in a practical iTTS, if the lookahead buffer is fed a short word, it may be preferable to delay its synthesis because  internal representation associated with it is likely to change when additional tokens become available. Also, it may be more useful to define the lookahead parameter in terms of future syllables rather than words. 
In addition,  perceptual evaluation shows that the dynamics between  encoder and  decoder are such that even if the encoder representation of an individual token changes slightly, the length of the encoder representation sequence will influence the way in which the decoder treats that token. 
We can conjecture that the decoder is regulating the duration of each segment with respect to sequence length. This will be addressed in future work by examining  attention weights the decoder uses when making predictions. 
Now that the importance of future context has been assessed, we also plan to work on context extension through prediction of future tokens using contextualized language models \cite{devlin-etal-2019-bert}.


\section{Acknowledgements}
This work was funded by the Multidisciplinary Institute in Artificial Intelligence MIAI@Grenoble-Alpes (ANR-19-P3IA-0003). The authors would like to thank Sylvain Gerber, Fanny Roche and Dave Green for fruitful discussions. 

\newpage
\bibliographystyle{IEEEtran}
\bibliography{refs}

\end{document}